\documentstyle[preprint,aps]{revtex}

\newcommand{\be}{\begin{equation}}
\newcommand{\ee}{\end{equation}}
\newcommand{\bea}{\begin{eqnarray}}
\newcommand{\eea}{\end{eqnarray}}
\newcommand{\bml}{\begin{mathletters}}
\newcommand{\eml}{\end{mathletters}}
\newcommand{\pa}{\partial}

\newcommand{\vx}{\vec{x}}
\newcommand{\vy}{\vec{y}}

\newcommand{\vk}{\vec{k}}

\newcommand{\e}{\epsilon}
\newcommand{\ve}{\varepsilon}

\begin{document}
\draft
\title{Gauge Dependence in the AdS/CFT Correspondence\cite{byline}}
\author{H. O. Girotti$^{a}$ and Victor O. Rivelles$^{b}$}
\address{$^{a}$Instituto de F\'{\i}sica,
Universidade Federal do Rio Grande do Sul \\ Caixa Postal 15051, 91501-970  -
Porto Alegre, RS, Brazil\\ E-mail: hgirotti@if.ufrgs.br \\
$^{b}$Instituto de F\'{\i}sica, Universidade de S\~ao Paulo\\
Caixa Postal 66318, 05315-970 - S\~ao Paulo, SP,
Brazil\\E-mail: rivelles@fma.if.usp.br }

\maketitle
\begin{abstract}
We consider the AdS space formulation of the classical dynamics deriving
from the St\"uckelberg Lagrangian. The on-shell action is shown to be
free of infrared singularities as the vector boson mass tends to zero. In
this limit the model becomes Maxwell theory formulated
in an arbitrary covariant gauge. Then we use the AdS/CFT
correspondence to compute the two-point correlation functions on the
boundary. It is shown that the gauge dependence concentrates on the
contact terms.

\end{abstract}
\pacs{PACS: 11.10.Kk, 11.25Mf \\
Keywords: AdS/CFT Correspondence, Maxwell theory, Gauge independence.}

\newpage
\narrowtext

As it is well known\cite{Mald}, Maldacena has conjectured that the
large $N$ limit of a certain conformal field theory (CFT) in a
d-dimensional space can be described 
by string/M-theory on $AdS_{d+1} \times K$, where $K$ is a
suitable compact space. A precise form to this conjecture has been
given in Refs.\cite{Gubser,Witt1} according to which

\be
\label{101}
Z_{AdS}[\phi_0]\,=\,\int_{\phi_0}\,{\cal D}\phi\,\exp(-
I[\phi])\,\equiv\,Z_{CFT}[\phi_0]\,=\, \langle \exp \left(\int_{\pa \Omega}\,
d^d x\,{\cal O}\phi_0\right)\rangle\,\,\,,
\ee

\noindent
where $\phi_0$ is the value taken by $\phi$ at the boundary. By assumption  
$\phi_0$ is also the external current coupling to the operator ${\cal O}$ in 
the boundary CFT. Thus the knowledge of the partition function in $AdS_{d+1}$
enables one to obtain the correlation functions of the boundary CFT in $d$
dimensions. The AdS/CFT correspondence has been studied for scalar
fields\cite{Witt1,Muck1,Freed,Bala,Minces2}, massive vector 
fields\cite{Muck2,Yi}, spinor fields\cite{Muck2,Henn,Ghez}, the 
Rarita-Schwinger field\cite{Volov,koshelev,viswa4}, classical 
gravity\cite{Liu,viswa5}, massive symmetric tensor fields \cite{polishchuk}, 
antisymmetric $p$-form fields\cite{frolov,l'yi}, type IIB string 
theory\cite{Banks,Chalm} and three dimensional field theories with 
Chern-Simons terms\cite{Riv1}. 

The AdS/CFT correspondence is an example of the holographic principle
\cite{holo} according to which a quantum theory with gravity must be
describable by a boundary theory. This raises questions on how the
detailed information of the theory in the bulk can be completely coded
in a lower dimensional theory at the border. In fact this mechanism is
still not well understood and several aspects of it have recently been
investigated. For instance, the holographic bound, establishing that the
boundary theory has only one bit of information per Planck area, manifests 
itself in the infrared-ultraviolet connection of the AdS/CFT correspondence
\cite{9805114}. Some situations involving superluminal oscillations and 
negative energy density have shown that there are hidden degrees of
freedom which store information but have no local energy density
\cite{pol}. On more conservative grounds, known relationships between field
theories in the bulk should emerge in the conformal theory at the
boundary. This has been verified explicitly for the case $AdS_3/CFT_2$. In 
fact, the well known equivalence between Maxwell-Chern-Simons
theory and the self-dual model in Minkowski space also holds in
$AdS_3$ and, correspondingly, both models have been shown to lead to the same 
conformal theory at the border \cite{Riv1}. 

Another aspect of the holographic principle is concerned with the
unphysical degrees of freedom of a gauge theory in the bulk. One expects 
that the AdS/CFT correspondence
respects gauge invariance in the sense that the CFT correlators are 
independent of the gauge choice made in the bulk. Since the
correlators of the corresponding conformal theory have conserved
sources they do not carry information about the longitudinal
modes of the gauge field. The conclusion would then be that there is no
information about the unphysical gauge degrees of freedom at the
border, which in turn would invalidate the AdS/CFT correspondence. In this 
work, we show that the gauge degrees of freedom do contribute but only to the
contact terms. To see how this come about, we shall consider an Abelian gauge
field and study the role played by the gauge dependent terms as far as the 
correlators at the border are concerned. 

Thus we shall be looking for a formulation of electrodynamics in an arbitrary
gauge when the space-time background is $AdS_{d+1}$. As in the case of
flat Minkowski space it will prove convenient to start from the St\"uckelberg
action 

\be
\label{102}
I_S\,=\,-\,\int\,d^{d+1}x\,{\sqrt g}\,\left[\,\frac{1}{4}\,F_{\alpha
    \beta}\,F^{\alpha \beta}\,+\,\frac{m^2}{2}\,
A_{\alpha}\,A^{\alpha}\,+\,\frac{1}{2a}\,\left(\nabla_{\alpha}
    A^{\alpha}\right)^2 \right]\,\,\,, 
\ee

\noindent
where $F_{\alpha \beta} = \pa_{\alpha}A_{\beta} - \pa_{\beta}A_{\alpha}$,
$\nabla_{\alpha}$ is the covariant derivative and $a$ is a real 
positive constant. Electrodynamics in an arbitrary
covariant gauge, specified by the constant $a$, is defined as the limit 
$m^2 \rightarrow 0$ of  St\"uckelberg theory. On the other hand the limit 
$a \rightarrow \infty$, while keeping $m^2 > 0$, results in the Proca theory.
The mass term in (\ref{102}) will help us to control the infrared divergent 
terms which will arise along the calculation. 

As usual we take the representation of $AdS_{d+1}$ in Poincar\'e coordinates 
which describes the half-space $x^0 > 0$, $x^i \in {\bf R^d}$ with the metric  

\be
\label{103}
g_{\mu \nu} = \frac{1}{(x^0)^2}\,\delta_{\mu \nu}\,\,\,.
\ee

\noindent
The Lagrange equations of motion arising from (\ref{102}) are found to read

\be
\label{1031}
\nabla_{\mu} F^{\mu \nu}\,+\,\frac{1}{a}\,\nabla^{\nu} \nabla_\mu
A^\mu\,-\,m^2 A^{\nu} \,=\,0\,\,\,.
\ee

Since we are going to solve the
equations of motion subjected to Dirichlet boundary conditions some care must
be exercised when applying the variational principle to the action (\ref{102}).
When we vary the action to obtain the equations of motion the boundary term 

\be 
\label{1032}
- \int d^d x \,\, \sqrt{g} \left( F^{0i} \delta A_i + \frac{1}{a}
  \nabla^\mu A_\mu \delta A^0 \right) \biggl. \biggr|_{x^0 = \e}
\ee
\noindent 
is generated. If the gauge fixing term is not present then the boundary 
conditions must be prescribed only for the spatial components of the potential
$A_i$. In the present case, however, all components of the potential must be
given at the border. Then no additional boundary terms are needed in the
original action to cancel the one coming from the variational principle.

The solving of the equations of motion is greatly simplified by the
decomposition of $A^{\mu}$ into a scalar field $\Phi$ and a vector
field $U^{\mu}$ 

\bml
\label{104}
\bea
\Phi\,&\equiv&\,\nabla_{\nu} A^{\nu}\,\,\,, \\
U^{\mu}\,&\equiv&\,A^{\mu}\,-\,\frac{1}{a m^2}\,\nabla^{\mu}\Phi\,\,\,. 
\eea
\eml

\noindent
These new fields satisfy, respectively, the equations of motion

\bml
\label{106}
\bea
&&\left( {\nabla^2} \,-\,a m^2\,\right)\Phi\,=\,0, \\
&&{\nabla_{\mu}} U^{\mu \nu}\,-\,m^2 U^{\nu}\,=\,0\,\,\,,
\eea
\eml

\noindent
where $U_{\mu\nu} \equiv \pa_{\mu}U_{\nu} - \pa_{\nu}U_{\mu}$. Clearly
$U^{\mu}$ is a Proca field with mass $m$ since $\nabla_{\mu}
U^{\mu}\,=\,0\,\,\,$.  

The solutions of the equations of motion (\ref{106}) 
converging at $x^0 \rightarrow \infty$ have already been presented in the 
literature\cite{Muck1,Muck2} and read, respectively,

\be
\label{109}
\Phi\,=\,(x^0)^{\frac{d}{2}}\,\int\,\frac{d^d k}{(2 \pi)^d}
\,e^{- i \vk \cdot \vx} \,\phi(\vk)\,K_{\alpha_a}(kx^0)\,\,\,, 
\ee

\bml
\label{110}
\bea
{\tilde U}_0(x)\,&=&\,(x^0)^{\frac{d}{2} + 1}\,\int\,\frac{d^d k}{(2 \pi)^d}
\,e^{- i \vk \cdot \vx}\,u_{0}(\vk)\,K_{{\tilde
\alpha}}(kx^0)\,\,\,,\label{mlett:a110} \\
{\tilde U}_i(x)\,&=&\,(x^0)^{\frac{d}{2}}\,\int\,\frac{d^d k}{(2
  \pi)^d}\,e^{- i \vk \cdot \vx}\, 
\left[u_i(\vk)\,K_{{\tilde \alpha}}(kx^0)\,+\,i\,u_0(\vk)
\,\frac{k_i}{k}\,x^0\,\,K_{{\tilde \alpha} + 1}(kx^0)\right]
\,\,\,,\label{mlett:b110}
\eea
\eml

\noindent
where $x \equiv (x^0, \vx)$, $k \equiv |\vk|$, $K$ is the modified Bessel
function and ${\tilde U}_{\mu} \equiv x^0 U_{\mu}$ 
are the components of $U$ with Lorentz indices\cite{Muck2}. Furthermore,

\bml
\label{112}
\bea
&&\alpha_a\,\equiv\,{\sqrt { am^2\,+\,\frac{d^2}{4}} }\,\,\,,
\label{mlett:a112}\\
&&{\tilde \alpha}\,\equiv\,{\sqrt{ \frac{(d - 2)^2}{4}\,+\,m^2}}\,\,\,.
\label{mlett:b112}
\eea
\eml

\noindent
We also recall that the Fourier transforms $u_0(\vk)$ and $u_i(\vk)$ are not 
all independent but related among themselves in order to secure the
fulfillment 
of $\nabla^\mu A_\mu=0$. It has been shown in \cite{Muck2} that  
$u_0\,=\,\frac{i k v}{\e}\,\frac{K_{{\tilde \alpha}}(k\e)}
{K_{{\tilde \alpha} + 1}(k\e)}$ and $u_i\,=\,v_i \,+\,v\, k_i$,
where $v\,=\,-\,\frac{k_i v_i}{k^2}\,\frac{k \e\, K_{{\tilde \alpha} +
    1}(k\e)} {\Sigma(k \e)}$,
\be
\label{115}
\Sigma(k \e)\,\equiv\,\left({\tilde \Delta}\,-\,1\right)
K_{{\tilde \alpha}}(k\e)\,+\,k\e\,K_{{\tilde \alpha} - 1}(k\e)\,\,\,,
\ee

\noindent
and ${\tilde \Delta}\,\equiv\,{\tilde \alpha}\,+\,\frac{d}{2}$. Here
$x^0 = \e>0$ specifies a near-boundary surface.  

As usual, we shall look for a bulk solution written in terms
of boundary field 
values specified at the near-boundary surface $x^0=\epsilon$, the limit 
$\epsilon \rightarrow 0$ \cite{Freed} being performed at the very end of the
calculations. In particular, by returning with (\ref{109}) and (\ref{110}) 
into (\ref{104}) one can determine the unknowns $\phi$ and $v_i$ in terms of
the values assumed by ${\tilde A}_{\mu}$ on the surface $x^0 = \e$. Thus,
one arrives at the following expressions for $\Phi$ and ${\tilde
U}_{\mu}$,

\bea
\label{117}
\Phi(x) &=& a m^2 \, (x^0)^{\frac{d}{2}} (\e)^{-\frac{d}{2}} \int
\frac{d^dk}{(2\pi)^d} e^{- i \vk \cdot \vx} \left[ i \epsilon k_j
\frac{K_{{\tilde \alpha}}(k \e)}{{\cal
D}(k\e)}  K_{\alpha_a}(kx^0)\,{\tilde A}_{\e,j}(\vk)  \right. \nonumber \\ 
&& \left. + \, \frac{\Sigma(k \e)}{{\cal D}(k\e)} 
K_{\alpha_a}(kx^0) \,{\tilde A}_{\e,0}(\vk)\right],
\eea 

\bea
\label{118}
{\tilde U}_0(x)\,&=& \,(x^0)^{\frac{d}{2} + 1}\,
\e^{-\left(\frac{d}{2} + 1\right)}
\,\int \frac{d^dk}{(2\pi)^d}\,e^{-i \vk \cdot \vx}\,
\left[- i \e k_j\,\left( \frac{1}{\Sigma (k\e)}\,-\,(k \e)^2\,
\frac{K_{{\tilde \alpha}}(k\e) K_{\alpha_a}(k\e)}{{\cal D}(k\e) \Sigma(k\e)}
\right) \right. \nonumber \\ 
&& \left. \times \,  K_{{\tilde \alpha}}(kx^0) {\tilde A}_{\e, j}(\vk)\,+\,
(k\e)^2\,\frac{K_{\alpha_a}(k\e)}{{\cal D}(k\e)} 
K_{{\tilde \alpha}}(kx^0) {\tilde A}_{\e, 0}(\vk) \right]\,\,\,,
\eea

\noindent
and

\bea
\label{119}
&{\tilde U}_i(x)&\,= \,(x^0)^{\frac{d}{2}}\,\e^{-\frac{d}{2}}
\,\int \frac{d^dk}{(2\pi)^d}\,e^{-i \vk \cdot \vx}\,
\left\{ \left[X_{ij}^{I} K_{{\tilde \alpha}}(kx^0) \,+\,kx^0\,
X_{ij}^{II} K_{{\tilde \alpha}+ 1}(kx^0) \right]{\tilde A}_{\e, j}(\vk) \right.
\nonumber\\
&&\left.  - \,\left[\left({\tilde \alpha} + 1 - \frac{d}{2}\right)\,i \e k_i\,
\frac{K_{\alpha_a}(k\e)}{{\cal D}(k\e)}
K_{{\tilde \alpha}}(kx^0) \,-i\,k k_i \, \e x^0\,
\frac{K_{\alpha_a}(k\e)}{{\cal D}(k\e)}\
 K_{{\tilde \alpha}+ 1}(kx^0) \right]{\tilde A}_{\e, 0}(\vk)
\right\}\,\,\,, 
\eea

\noindent
where

\bml
\label{122}
\bea
&& X_{ij}^{I}\,=\,\frac{1}{K_{{\tilde \alpha}}(k\e)} \left[\delta_{ij} -
(k \e) \frac{k_i k_j}{k^2} \frac{K_{{\tilde \alpha} + 1}(k\e)}{\Sigma (k\e)}
\right] + \left({\tilde \alpha} + 1 - \frac{d}{2}\right)\,(\e)^2 k_i k_j\,
\frac{K_{{\tilde \alpha}}(k\e) K_{\alpha_a}(k\e)}{{\cal D}(k\e) \Sigma(k\e)}
,\label{mlett:a122}\\
&& X_{ij}^{II}\,=\,\frac{k_i k_j}{k^2} \frac{1}{\Sigma (k\e)}\,-\,(\e)^2 k_i k_j\,
\frac{K_{{\tilde \alpha}}(k\e) K_{\alpha_a}(k\e)}{{\cal D}(k\e) \Sigma(k\e)}
\,\,\,,\label{mlett:b122}
\eea
\eml

\be
\label{124}
{\cal D}(k\e)\,\equiv\,(k \e)^2\,K_{{\tilde \alpha}}(k\e) K_{\alpha_a}(k\e)\,
+\,\Sigma(k\e) \,\Lambda(k\e)\,\,\,,
\ee

\noindent
and

\be
\label{125}
\Lambda(k\e)\,\equiv\,\left( \frac{d}{2} - \alpha_a \right)K_{\alpha_a}(k\e)
\,-\,k\e\,K_{\alpha_a - 1}(k\e)\,\,\,.
\ee

\noindent
Furthermore, $
{\tilde A}_{\e, \mu}(\vk)\,=\,\int\, d^dx\,e^{i \vk \cdot \vx}\,
{\tilde A}_{\e, \mu}(\vx)$, where $
{\tilde A}_{\e, \mu}(\vx)\,\equiv\,{\tilde A}_{\mu}(x^0=\e, \vx)$.
Notice that when $a \rightarrow \infty$  we reproduce the results
found in Ref.\cite{Muck2} for the Proca field. 

In this paper we are interested in Maxwell theory formulated in an
arbitrary covariant gauge. Therefore we must investigate the limit 
$m^2 \rightarrow 0$ of (\ref{117}--\ref{119}). From Eqs.(\ref{mlett:a112}) and
(\ref{mlett:b112}) one finds, respectively, that the expansions

\bml
\label{132}
\bea
&&{\tilde \alpha}\,=\,\frac{d}{2}\, -\, 1\, +\, \frac{ m^2}{d-2}\,\,\,,
\label{mlett:a132}\\
&&\alpha_a \,=\,\frac{d}{2}\,+\, \frac{a m^2}{d}\,\,\,,\label{mlett:b132}
\eea
\eml

\noindent
are valid up to terms of order $m^2$. Then from Eqs.(\ref{115}), (\ref{124}) 
and (\ref{125}) it follows that ${\cal D}(k\e)$ vanishes as $m^2$ goes to zero.
In fact, it is straightforward to show that

\bea
\label{134}
\lim_{m^2 \rightarrow 0}{\cal D}(k\e) &\rightarrow & m^2\,\left\{ (k\e)^2\,
\frac{(a-1)d-2a}{d(d-2)}  \left[K_{\frac{d}{2}-1}(k\e) 
\frac{\pa K_{\rho}(k\e)}{\pa \rho}\biggl. \biggr|_{\rho=\frac{d}{2}} - 
K_{\frac{d}{2}}(k\e)
\frac{\pa K_{\rho}(k\e)}{\pa \rho}\biggl. \biggr|_{\rho=\frac{d}{2}-1}\right]
\right. \nonumber\\ 
&& \left.+ 
\frac{k\e}{d-2} \,K_{\frac{d}{2}-1}(k\e)K_{\frac{d}{2}-1}(k\e)\,-\,
\frac{a\,k\e}{d} \,K_{\frac{d}{2}}(k\e)K_{\frac{d}{2}}(k\e) \right\} \,\,\,.
\eea

\noindent
On the other hand, (\ref{mlett:a132}) implies that $\lim_{m^2
  \rightarrow 0} \left( {\tilde \alpha} + 1 - \frac{d}{2}
  \right)\,=\,\frac{m^2}{d-2}$, which also vanishes as $m^2
  \rightarrow 0$. 

\noindent  
This implies that $\Phi/am^2$ and $U_{\mu}$ develop infrared divergences. The 
question is now whether $A_{\mu}$ is well defined in the zero mass limit. To 
investigate this point we go back with Eqs.(\ref{117}-\ref{119}) into 
Eq.(\ref{104}) obtaining

\bea
\label{136}
&&{\tilde A}_0(x)\,=\,(x^0)^{\frac{d}{2}}\,(\e)^{- \frac{d}{2}}\,\int\,
\frac{d^dk}{(2 \pi)^d}\,e^{-i \vk \cdot \vx}
\left\{\left[
-\,\frac{x^0}{\e}\,\frac{i \e k_j}{\Sigma(k\e)}\,K_{{\tilde
\alpha}}(kx^0) 
+\,i \frac{d}{2}\,
\ \e k_j\,\frac{K_{{\tilde \alpha}}(k \e)}{{\cal
D}(k\e)} \, K_{\alpha_a}(kx^0)\,
\right. \right. \nonumber\\
&&  \left. \left. + i \,\e x^0 \, k_j\,(k\e)^2 \frac{K_{{\tilde
\alpha}}(k\e) K_{\alpha_a}(k\e)}{{\cal D}(k\e) \Sigma(k\e)} K_{{\tilde
\alpha}}(kx^0) \,
+\, \,i k_j\,k \, \e x^0 \, \frac{K_{{\tilde \alpha}}(k
      \e)}{{\cal D}(k\e)} \, \frac{\pa 
K_{\alpha_a}(kx^0)}{\pa (kx^0)}\,\right]{\tilde A}_{\e, j}(\vk)
 \right.  \nonumber \\
&& \left.  +  \left[\frac{d}{2}\,
\frac{\Sigma(k \e)}{{\cal D}(k\e)}
\, K_{\alpha_a}(kx^0)\,
+\, kx^0\, \frac{\Sigma(k \e)}{{\cal D}(k\e)} \frac{\pa
  K_{\alpha_a}(kx^0)}{\pa (kx^0)}\,+\,  
k^2 \, \e x^0 \,\frac{K_{\alpha_a}(k\e)}{{\cal D}(k\e)}
\, K_{{\tilde\alpha}}(kx^0) \right]{\tilde A}_{\e, 0}(\vk)\right\}\,\,\,
\eea

\noindent
and  

\bea
\label{137}
&&{\tilde A}_i(x)\,=\,(x^0)^{\frac{d}{2}}\,(\e)^{- \frac{d}{2}}\,\int\,
\frac{d^dk}{(2 \pi)^d}\,e^{-i \vk \cdot \vx}
\left\{\left[X_{ij}^{I}\,K_{{\tilde \alpha}}(kx^0)\,
+\,kx^0\,\frac{k_ik_j}{k^2} \frac{1}{\Sigma(k\e)}\,
K_{{\tilde \alpha} + 1}(kx^0) \right. \right.
\nonumber\\
&& \left. \left.-\,kx^0\,(\e)^2 k_i k_j \frac{K_{{\tilde \alpha}}
(k\e) K_{\alpha_a}(k\e)} {{\cal D}(k\e) \Sigma(k\e)}\,
K_{{\tilde \alpha} + 1}(kx^0)  + \e x^0 k_i
\, k_j\,\frac{K_{{\tilde \alpha}}(k \e)}{{\cal D}(k\e)}
\,
K_{\alpha_a}(kx^0)\right] {\tilde A}_{\e,j}(\vk) \right.\nonumber\\
&&\left. -  \left[
\left({\tilde \alpha} + 1 - \frac{d}{2}\right)\,i \e k_i\,
\frac{K_{\alpha_a}(k\e)}{{\cal D}(k\e)}\,
\,K_{{\tilde \alpha}}(kx^0)\,- i \,k k_i\,\e x^0 \,
\frac{K_{\alpha_a}(k\e)}{{\cal D}(k\e)}
\,
K_{{\tilde \alpha} + 1}(kx^0)\, \right. \right. \nonumber \\
&& \left. \left. +\,i x^0 k_i\,\frac{\Sigma(k \e)}{{\cal D}(k\e)}
\,K_{\alpha_a}(kx^0) \right] {\tilde A}_{\e,0}(\vk) \right\}\,\,\,.
\eea

\noindent
The three terms in the second line of (\ref{136}) are, individually, infrared
divergent. However, by using

\bml
\label{139}
\bea
&&K_{\alpha_a}\,=\,K_{\frac{d}{2}}\,+\,{\it O}(m^2)\,\,\,,
\label{mlett:a139}\\
&&K_{{\tilde \alpha} + 1}\,=\,K_{\frac{d}{2}}\,+\,{\it O}(m^2)\,\,\,,
\label{mlett:b139}\\
&&\Sigma(k\e)\,=\,k\e\,K_{\frac{d}{2}}(k\e)\,+\,{\it O}(m^2)\,\,\,,
\label{mlett:c139}\\
&&\Lambda(k\e)\,=\,-\,k\e\,K_{\frac{d}{2} - 1}(k\e)\,+\,{\it O}(m^2)\,\,\,,
\label{mlett:d139}
\eea
\eml
 
\noindent
one finds that the divergent pieces cancel among themselves. 
Through a similar analysis we show that the third line in Eq.(\ref{136}) and 
the second and third lines in Eq.(\ref{137}) define functions of $m^2$ which 
are regular at $m^2 = 0$. To summarize, $A_{\mu}(x)$ is indeed an analytic
function of $m^2$ in the vicinity of $m^2 = 0$.

We turn next into application of the AdS/CFT correspondence to compute the
two-point correlation function $<{\cal O}_{\mu}(\vx) {\cal O}_{\nu}(\vy)>$
of the
boundary CFT. The dominant term in the path integral in Eq.(\ref{101}) is the
exponential of the classical action evaluated on-shell. After using
Eqs.(\ref{102}) and (\ref{1031})  one finds that

\bea
\label{140}
I_S\,&=&\,\frac{1}{2}\,\int_{\pa \Omega}\,d^dx\,\e^{-d}\,\left\{ {\tilde
A}_{\e,i}(\vx) \left[-\,{\tilde A}_{\e,i}(\vx)\,+\,\e\,{\tilde F}_{\e,0i}(\vx)
\right] \right. \nonumber\\
&& \left. +\,\frac{1}{a}\,{\tilde A}_{\e,0}(\vx)\,
\left[ -d\,{\tilde A}_{\e,0}(\vx)\,+\,\e\,\left( \frac{\pa{\tilde A}_0(x)}
{\pa x^0}\biggl. \biggr|_{x^0 = \e}\,+\,\pa_i {\tilde
A}_{\e,i}\right)\right]\right\} \,\,\,,
\eea

\noindent
where ${\tilde F}_{0i}(x) \equiv \pa_0{\tilde A}_i(x) - \pa_i{\tilde A}_0(x)$
and $\pa\Omega$ is the near-boundary surface $x^0 = \e$. The explicit
dependence of $I_S$ on the gauge parameter $a$ should be noticed. Non trivial
field theory correlators may only arise from those terms containing
${\tilde F}_{\e,0i}(\vx)$ and $\pa{\tilde A}_0(x)/
\pa x^0|_{x^0 = \e}$. We shall therefore concentrate our attention on these 
objects.

>From (\ref{136}) and (\ref{137}) and after setting $x^0 = \e$ one obtains 

\bea
\label{141}
&&{\tilde F}_{\e,0i}(\vx)\,=\,\left(\frac{d}{2} - {\tilde \alpha}\right)\,
\frac{1}{\e} \,{\tilde A}_{\e,i}(\vx)\nonumber\\
&& +\,\int \frac{d^dk}{(2\pi)^d}\,
e^{-i \vk \cdot \vx}\,\left[\frac{k K_{{ \tilde\alpha} - 1}}
{K_{{ \tilde\alpha}}}\,
\left(- \delta_{ij}\,+\,\frac{k_i k_j}{k^2}\,k\e\,
\frac{K_{{ \tilde\alpha} + 1}}
{\Sigma}\right)\right] {\tilde A}_{\e,j}(\vk)\nonumber\\
&& + \left({\tilde \Delta} - 1 \right)\,\left( {\tilde \alpha} + 1
- \frac{d}{2}\right)\,\int \frac{d^dk}{(2\pi)^d}\,e^{-i \vk \cdot \vx}\,
\left[\e k_i k_j \frac{K_{\alpha_a} K_{{\tilde \alpha}}^2}{\Sigma {\cal D}}\,
{\tilde A}_{\ve,j}(\vk)\,-\,i k_i\,\frac{K_{\alpha_a} K_{{\tilde \alpha}}}
{{\cal D}}\,{\tilde A}_{\e,0}(\vk) \right],
\eea 

\noindent
where the argument $ k\e$ of the modified Bessel functions has been omitted in
order to simplify the notation. The first two lines in the right hand side of
(\ref{141}) survive in the limit $a \rightarrow \infty$ and reproduce, as it
must be the case, the result for the Proca field \cite{Muck2}.
Furthermore the dangerous infrared behavior of ${\cal D}$, showing up in the
denominator of the last line of (\ref{141}), is again canceled by the factor
${\tilde \alpha}+1-d/2$ in the corresponding numerator, leaving us with an 
overall function ${\tilde F}_{\e,0i}(\vx)$ regular at $m^2 = 0$.  

As for $\pa{\tilde A}_0(x)/\pa x^0|_{x^0 = \e}$ Eq.(\ref{136}) leads to

\bea
\label{142}
&&\frac{\pa{\tilde A}_0(x)}{\pa x^0}\biggl. \biggr|_{x^0 = \e}\,=\,
\left(\frac{d}{2}\, +\, 1\right)\,\frac{1}{\e}\,{\tilde A}_{\e,0}(\vx)\,+\,
\frac{1}{\e}\,\int\,\frac{d^dk}{(2\pi)^d}\,e^{-i \vk \cdot \vx}\nonumber\\
&&\times \left\{ \left[ 1\,+\,\left(\alpha_a - \frac{d}{2}\right)^2
\frac{K_{\alpha_a} K_{{\tilde \alpha}}}{\cal D}\right]\,{\tilde A}_{\e,j}(\vk)
\,+\,\left[\left(\frac{d}{2}\,-\,1\right)\,
+\,\left(\alpha_a^2\,-\,\frac{d^2}{4}\right)\,\frac{\Sigma K_{\alpha_a}}{\cal
D}\right]\,{\tilde A}_{\e,0}(\vk)\right\}. 
\eea

\noindent
Since $(\alpha_a - \frac{d}{2})$ and ${\cal D}$ are ${\it O}(m^2)$ the second
term in the first bracket of Eq.(\ref{142}) is ${\it O}(m^2)$ and, therefore,
it drops out in the zero mass limit. For the same reasons the term involving 
$1/{\cal D}$ in the second bracket of (\ref{142}) is regular in $m^2=0$ which
renders $\pa{\tilde A}_0(x)/\pa x^0|_{x^0 = \e}$ free of infrared divergences.
  
We shall next determine the contributions of each term 
in (\ref{141}) and (\ref{142}) to the action in (\ref{140}) and, therefore, to
the correlators $<{\cal O}_{\mu}(\vx) {\cal O}_{\nu}(\vy)>$. We shall do this
for $m^2 > 0$, the limit $m^2 \rightarrow 0$ being taken at the very end of the
calculations. 

Clearly the first line in the right hand side of (\ref{141})
contributes with a contact term which is independent of $a$. 
The second line in the right hand side of (\ref{141}) gives the usual
result for the two point correlator after taking the usual
prescription $A_{0,\mu}(\vx)\,=\,\lim_{\e \rightarrow 0}\,
\e^{{\tilde \Delta} - d}\,{\tilde A}_{\e,\mu}(\vx)$.
Notice that ${\tilde \Delta} - d$ is gauge independent. Hence, at the limit 
$a \rightarrow \infty$ we have  
$U_{0,\mu}(\vx)\,=\,\lim_{\e \rightarrow 0}\,
\e^{{\tilde \Delta} - d}\,{\tilde U}_{\e,\mu}(\vx)$,
which, as expected, reproduces the normalization
prescription for the Proca field\cite{Muck2}. Then by taking into
account the above normalization and using the AdS/CFT 
correspondence one reads off the two-point correlators

\be
\label{151}
<{\cal O}_i(\vx) {\cal O}_j(\vy)>\biggl. \biggr|_{m^2=0}\,=\,
\frac{\Gamma(d)}{\pi^{d/2} \Gamma(\frac{d-2}{2})} 
\times\, \frac{1}{|\vx - \vy|^{2d - 2}} \left[\delta_{ij}\,-\,2 
\frac{(x_i - y_i)(x_j - y_j)}{|\vx - \vy|^2}\right]\,\,\,,
\ee

\noindent
in agreement with the results already obtained for this object in the case of
the Abelian gauge field\cite{Witt1,Freed}.

The third line in the right hand side of Eq.(\ref{141}) only
makes trivial contributions to the two-point CFT correlator. Thus if 
$<{\cal O}_{\mu}(\vx) {\cal O}_{\nu}(\vy)>$ contains at all a nontrivial gauge
dependent piece it can only originate in the last term of the right
hand side of Eq.(\ref{142}) which, as we already said, survives in the limit 
$m^2 \rightarrow 0$.  
However, an straightforward calculation suffices to show that all potentially
dangerous powers of $\e$ cancel out among themselves 
and the just mentioned term does not contribute to 
$<{\cal O}_{\mu}(\vx){\cal O}_{\nu}(\vy)>$ in
the limit $\e \rightarrow 0$.

We then conclude that the gauge dependence concentrates on the
contact terms while the non-trivial part of the boundary conformal theory 
correlators turns out to be that already found by working in a
completely fixed gauge\cite{Witt1,Freed} and displayed in
Eq.(\ref{151}). Another important feature is that although we have
fixed all components of the potential at the border the pieces
containing $\tilde{A}_{\epsilon,0}$ give only contact terms and the only
non-trivial pieces are those containing $\tilde{A}_{\epsilon,i}$. Therefore
the boundary theory still retains information on the gauge degrees of
freedom of the bulk theory. This then lends further support to the
holographic principle. Our result confirms the expectation that the
AdS/CFT correspondence respects gauge invariance and that the
information about the unphysical degrees of freedom is not lost in the
border. This shows the importance of the contact terms which are
usually ignored.


\newpage

\begin{references}

\bibitem[*]{byline} Supported in part by Conselho Nacional de Desenvolvimento
Cient\'{\i}fico e Tecnol\'ogico (CNPq) and Funda\c c\~ao de
Amparo \`a Pesquisa do Estado de S\~ao Paulo (FAPESP). 

\bibitem {Mald} J. Maldacena, ``The Large N Limit of Superconformal Field
Theories and Supergravity'', hep-th/9711200, Adv. Theor. Math. Phys. {\bf 2}
231 (1998). 

\bibitem {Gubser} S. Gubser, I. Klebanov and A. Polyakov, ``Gauge Theory
Correlators from Non-Critical String Theory'', hep-th/9802109, Phys. Lett. {\bf
428} 105 (1998).

\bibitem {Witt1} E. Witten, ``Anti-de Sitter Space and Holography'',
hep-th/9802150, Adv. Theor. Math. Phys. {\bf 2} 253 (1998). 

\bibitem {Muck1} W. M\"uck and K. S. Viswanathan, ``Conformal Field Theory
Correlators from Classical Scalar Field Theory on $AdS_{d+1}$'',
hep-th/9804035, Phys. Rev. {\bf D58} 41901 (1998).

\bibitem {Freed} D. Z. Freedman, S. D. Mathur, A. Matusis and L. Rastelli,
``Correlation Functions in the $CFT_d/AdS_{d+1}$ Correspondence'',
hep-th/9804058, Nucl. Phys. {\bf B546} 96 (1999).

\bibitem {Bala} V. Balasubramanian, P. Kraus and A. Lawrence, ``Bulk vs.
Boundary Dynamics in Anti-de Sitter Spacetime'', hep-th/9805171, Phys.
Rev. {\bf D59} 046003 (1999).

\bibitem {Minces2} P. Minces and V. O. Rivelles, ``Scalar Field Theory
  in the AdS/CFT Correspondence Revisited'', hep-th/9907079.

\bibitem {Muck2} W. M\"uck and K. S. Viswanathan, ``Conformal Field Theory
Correlators from Classical Field Theory on Anti-de Sitter Space II. Vector and
Spinor Fields'', hep-th9805145, Phys. Rev. {\bf D58} 106006 (1998).

\bibitem {Yi} W. S. l'Yi, ``Coordinate-Space Holographic Projection of
  Fields and an Application to Massive Vector Fields'', hep-th/9808051.

\bibitem {Henn} M. Henningson and K. Sfetsos, ``Spinors and the AdS/CFT
Correspondence'', hep-th/9803251, Phys. Lett. {\bf B431} 63 (1998).

\bibitem {Ghez} A. Ghezelbash, K. K. Kaviani, S. Parvizi and A. Fatollahi,
``Interacting Spinors-Scalars and AdS/CFT Correspondence'', hep-th/9805162,
Phys. Lett. {\bf B435} 291 (1998).

\bibitem {Volov} A. Volovich, ``Rarita-Schwinger Field in the AdS/CFT
Correspondence'', hep-th/9809009, J. High. En. Phys. {\bf 9809} 22 (1998).

\bibitem{koshelev} A. Koshelev and O. Rytchkov, ``Note on the Massive
Rarita-Schwinger Field in the AdS/CFT correspondence'', hep-th/9812238,
Phys. Lett. {\bf B450} 368 (1999).

\bibitem{viswa4} P. Matlock and K. S. Viswanathan, ``The AdS/CFT
Correspondence for the Massive Rarita-Schwinger Field'', hep-th/9906077.

\bibitem {Liu} H. Liu and A. Tseytlin, ``D=4 Super Yang-Mills, D=5 Gauged
Supergravity and D=4 Conformal Supergravity'', hep-th/9804083,
Nucl. Phys. {\bf B533} 88 (1998).

\bibitem{viswa5} W. M\"uck and K. S. Viswanathan, ``The Graviton
in the AdS-CFT Correspondence: Solution Via the Dirichlet Boundary Value
Problem'', hep-th/9810151.

\bibitem{polishchuk} A. Polishchuk, ``Massive Symmetric Tensor Field on
AdS'', hep-th/9905048, J. High En. Phys. {\bf 9907} 007 (1999).

\bibitem{frolov} G. Arutyunov and S. Frolov, ``Antisymmetric Tensor Field
on $AdS_5$'', hep-th/9807046, Phys. Lett. {\bf B441} 173 (1998).

\bibitem{l'yi} W. l'Yi, ``Correlators of Currents Corresponding to the
Massive $p$-form Fields in AdS/CFT Correspondence'', hep-th/9809132,
Phys. Lett. {\bf B448} 218 (1999).

\bibitem {Banks} T. Banks and M. Green, ``Non-perturbative Effects in
AdS5$^{\ast}$S5 String Theory and d=4 SUSY Yang-Mills'', hep-th/9804170, J.
High. En. Phys. {\bf 9805} 002 (1998).

\bibitem {Chalm} G. Chalmers, H. Nastase, K. Schalm and R. Siebelink,
``R-Current Correlators in N=4 Super Yang-Mills Theory from Anti-de Sitter
Supergravity'', hep-th/9805105.

\bibitem {Riv1} P. Minces and V. O. Rivelles, ``Chern-Simons Theories in the
AdS/CFT Correspondence'', hep-th/9902123, Phys. Lett. {\bf B455} 147 (1999). 

\bibitem{holo} G. 'tHooft, ``Dimensional Reduction in Quantum
  Gravity'', gr-qc/9310026; L. Susskind, ``The World as a Hologram'', 
  hep-th/9409089, J. Math. Phys. {\bf 36} 6377 (1995).

\bibitem{9805114} L. Susskind and E. Witten, ``The Holographic Bound
  in Anti-de Sitter Space'', hep-th/9805114.

\bibitem{pol} J. Polchinski, L. Susskind and N. Toumbas, ``Negative
Energy, Superluminosity and Holography'', hep-th/9903228. 

\end{references}
\end{document}